\documentclass[manuscript]{aastex}
\usepackage{epstopdf,natbib}
\usepackage{fullpage}
\usepackage{epstopdf,natbib}
\usepackage{graphicx} 
\usepackage{wrapfig}
\usepackage{amsmath}
\usepackage[caption=false]{subfig}

\shortauthors{Nugent et al.}
\begin{document}

\title{Observed asteroid surface area in the thermal infrared}

\author{C. R. Nugent\altaffilmark{1}, 
A. Mainzer\altaffilmark{2}, 
J. Masiero\altaffilmark{2}, 
E. L. Wright\altaffilmark{4}, 
J. Bauer\altaffilmark{2}, 
T. Grav\altaffilmark{3}, 
E. A. Kramer\altaffilmark{2}, and
S. Sonnett\altaffilmark{2}
}

\altaffiltext{1}{Infrared Processing and Analysis Center, California Institute of Technology, Pasadena, CA 
91125, USA}
\altaffiltext{2}{Jet Propulsion Laboratory, California Institute of Technology, Pasadena, CA 91109, USA}
\altaffiltext{3}{Planetary Science Institute, Tucson, AZ}
\altaffiltext{4}{Department of Physics and Astronomy, University of California, Los Angeles, CA 90095, USA}

\begin{abstract} 

The rapid accumulation of thermal infrared observations and shape models of asteroids has led to increased interest in thermophysical modeling. Most of these infrared observations are unresolved. We consider what fraction of an asteroid's surface area contributes the bulk of the emitted thermal flux for two model asteroids of different shapes over a range of thermal parameters. The resulting observed surface in the infrared is generally more fragmented than the area observed in visible wavelengths, indicating high sensitivity to shape. For objects with low values of the thermal parameter, small fractions of the surface contribute the majority of thermally emitted flux. Calculating observed areas could enable the production of spatially-resolved thermal inertia maps from non-resolved observations of asteroids.

\end{abstract}

\section{Introduction}

The availability of three-dimensional asteroid shapes allows for computation of observed surface area. This computation is routinely done for objects observed by radar \citep[e.g.][Figure 7]{2010BrozovicMithra}, as well as resolved spacecraft observations of asteroids \citep[e.g.][Figure 2]{11SteinsLeyrat}. However, less consideration has been given to the fraction of surface observed in the infrared when the asteroid itself is not resolved. 

Unresolved observations of asteroids are an integrated sum of flux from the surface of an asteroid visible to the observer. The power radiated from a black body is proportional to the fourth power of temperature; consequently, a small fraction of surface area can contribute the majority of power emitted. Therefore, infrared observations of asteroids are often sensitive to only a small patch of surface area. Additionally, the size of this patch is dependent on the average surface thermal inertia and the wavelengths used for observation. 

With the rapid accumulation of thermal-IR observations of small bodies, the improvement of modeling techniques, and the increase in computational power, thermophysical modeling, which combines infrared observations with asteroid shapes to determine surface thermal inertia, is a growing field of interest \citep[e.g.][]{Muller2005AA, Wright07, 11SteinsLeyrat, Coradini11, Matter11, 11RozitisGreen,Delbo.2011a, 13Keihm, 14EmeryBennu, 14AliLagoaEV5, 2015Koren, DelboAsteroidsIV}. Thermal modeling has a long heritage \citep[e.g.][]{Brown85,Lebofsky86,Spencer89,Lagerros96a, Lagerros96b, Lagerros97, 
Lagerros98, Tedesco02}. Recent growth is partly motivated by a desire to employ the wealth of infrared measurements of asteroids made by surveys such as IRAS \citep{Tedesco02}, NEOWISE \citep{Wright10WISE,14MainzerRestart}, and AKARI \citep{13Usui} which have observed hundreds of thousands of asteroids as point sources. 

Previous thermal models, such as the Near-Earth Asteroid Thermal Model \citep[NEATM,][]{Harris98}, are effective for measuring diameters and albedos but include many assumptions, including a spherical, non-rotating body with zero emission from the night side of the object. The Fast-Rotating Model (FRM) assumes a spherical object with latitudinal bands of uniform temperature and $0^{\circ}$ obliquity. Thermophysical models aim to more accurately model asteroids by computing heat transport and including additional data such as radar-derived shapes and spin poles, they are separate from and more computationally intensive than NEATM or FRM. Thermophysical models generally incorporate the effects of small-scale surface roughness.

The influence of shape on thermal modeling has also been investigated. \citet[see Figure 5]{Delbo02} compared the temperature distribution for (6489) Golevka with the temperature distribution produced by a simplified model. \citet{15Hanus} investigated how uncertainty in shape model and spin orientation impact thermophysical modeling results.

To facilitate thermophysical modeling of unresolved point sources, we quantified the observed area for two modeled asteroid shapes over a range of parameters. These results offer insight into the difference between the part of the surface that is visible, geometrically, to the telescope, and the part of the surface that contributes to the observed infrared flux. 

\section{Methods}

A useful metric for describing the thermal environment of a rotating body is the unit-less thermal parameter, $\Theta$ \citep[e.g.][]{1971WinterKrupp, Spencer89}. It is defined as
\begin{equation} 
\Theta = \frac{\sqrt{K \rho C \omega}}{\epsilon \sigma T^3}
\end{equation}
Where $K$ is the thermal conductivity, $\rho$ is the density, $C$ is the heat capacity, $\omega$ is the body's angular rotation rate, $\epsilon$ is the emissivity, $\sigma$ is the Stefan-Boltzmann constant, and $T$ is the sub-solar 
temperature of the surface. This can be rewritten in terms of thermal inertia $\Gamma$ as 
\begin{equation} 
\Theta = \frac{ \Gamma \sqrt{ \omega}}{\epsilon \sigma T^3}
\end{equation}
As objects move farther away from the sun, $T$ decreases, causing $\Theta$ to increase. 
The  $\Theta$ of an object is independent of diameter.
When the surface of the body is in equilibrium with the incident radiation, $\Theta = 0$. Objects with low $\Theta$ are often referred to as ``slow rotators", as heat is conducted and re-emitted quickly relative to the rotation rate. A body with uniform surface temperature would have $\Theta$ approaching infinity, and objects with high $\Theta$ are termed ``fast rotators" (Spencer et al., 1989).  In other words, $\Theta$ is the ratio of an effective thermal conductivity due to conduction to the thermal emission due to radiation. 

Another relevant parameter is the unit-less x value, where 
\begin{equation}
x=\frac{h \nu}{kT}
\end{equation}
and $k$ is the Boltzmann constant, $\nu$ is the frequency, and $h$ is Planck's constant. Although the emitted power from the surface is proportional to the fourth power of temperature, the x value expresses the emitted brightness at a given band. Individual bands probing the Rayleigh-Jeans side of the blackbody will be less sensitive to temperature variations than bands probing the Wein's approximation side of the curve. The x value is calculated for the resulting models, with $T$ taken to be the peak temperature (the highest temperature on the surface) and $\nu$ corresponding to the frequency of the NEOWISE $12$ $\mu m$ band.

We modeled two asteroid shapes for a range of $\Theta$, and calculated the surface areas that contributed $68\%$ and $95\%$ of the observed flux. One shape was a sphere with 800 triangular facets, generated with code based on that of \citet{KaasalainenIII} and  implemented as described in \citet{Mainzer2011b}. The other shape was a radar-derived 1996-facet shape of (4486) Mithra derived assuming prograde rotation from \citet{2010BrozovicMithra}. This calculation is representative of dog bone or dumbbell-shaped asteroids, and temperature is calculated at 1 AU from the sun and with $0^{\circ}$ obliquity. Therefore, the results shown here are not reflective of the actual temperatures on 4486 Mithra, which has a non-zero obliquity, a 0.6 orbital eccentricity, and has an orbital semimajor axis of 2.2 AU. Calculations assumed a visible geometric albedo of 0.2, and zero obliquity. Heat capacity was set at 500 J kg$^{-1}$ K$^{-1}$, thermal conductivity at 0.1 W m$^{-1}$ K$^{-1}$, and density at 2000 kg m$^{-3}$, roughly following values measured by \citet{OpeilTherCon} for ordinary and carbonaceous chondrites. 

The effects of small-scale surface roughness decrease as $\Theta$ increases \citep{Lagerros97}, and depend on wavelength \citep{2014PASJMuller}. We modeled roughness via the effective infrared emissivity, following the statistical treatment of roughness described in  \citet{11SteinsLeyrat} which examined the main-belt asteroid (2867) Steins. There the authors invoke the self-heating parameter $\xi$ of \citet{Lagerros97} and employ the visible phase curves of Steins to determine an effective emissivity, $\epsilon_{IR_{eff}}=0.73 \pm 0.02$, for that object. In this work, the effects of roughness were modeled by testing $\epsilon_{IR_{eff}}=0.7$ and $0.5$, as well as a case without roughness with $\epsilon_{IR}=0.9$. Emissivity of 0.9 follows typical values found for meteorites \citep{16Maturilli}.
 
To change $\Theta$, the rotation period was varied between 0.001 and 1000.0 hours in logarithmic steps, while $\Gamma$ was held constant. However, this choice was arbitrary; identical results could be obtained by varying $\Gamma$ while maintaining a fixed rotation period. Resulting $\Theta$ ranged between $4.2 \times 10^2$ and $4.2 \times 10^{-1}$. The spherical asteroid had a diameter of 100 meters, though we note again that since $\Theta$ is independent of diameter, the results do not depend on diameter and are valid for the sizes of known asteroids. (The usefulness of $\Theta$ breaks down at pathological extremes, where the thermal depth wave is on the order of the size of the body, such as for cometary dust grains). The asteroid was placed at 1 AU from the sun, on a circular orbit. Observed fraction was calculated both at opposition and at $90^\circ$ phase angle. Observed fraction was taken to be the number of facets that contributed $68\%$ of the flux divided by half the total number of facets.

To calculate heat transport through the body, we employed SINDA, a three-dimensional heat transfer code used across a variety of disciplines, including manufacturing engineering and spacecraft design \citep{CRTechwebsite}. The object underwent a warm up phase to allow interior temperatures to equilibrate before final surface temperatures were calculated; equilibration was defined as when the total flux of the asteroid changed by less than $1\%$. Self-shadowing and self-heating were both calculated, assuming Lambertian reflection. From the resultant temperature maps, thermal emission and reflected sunlight was calculated for each facet.

\section{Results}

\begin{figure}[t!]
    \subfloat[$\Theta = 4.2 \times 10^2$, x = 4.5]{
    \includegraphics[width=0.31\textwidth]{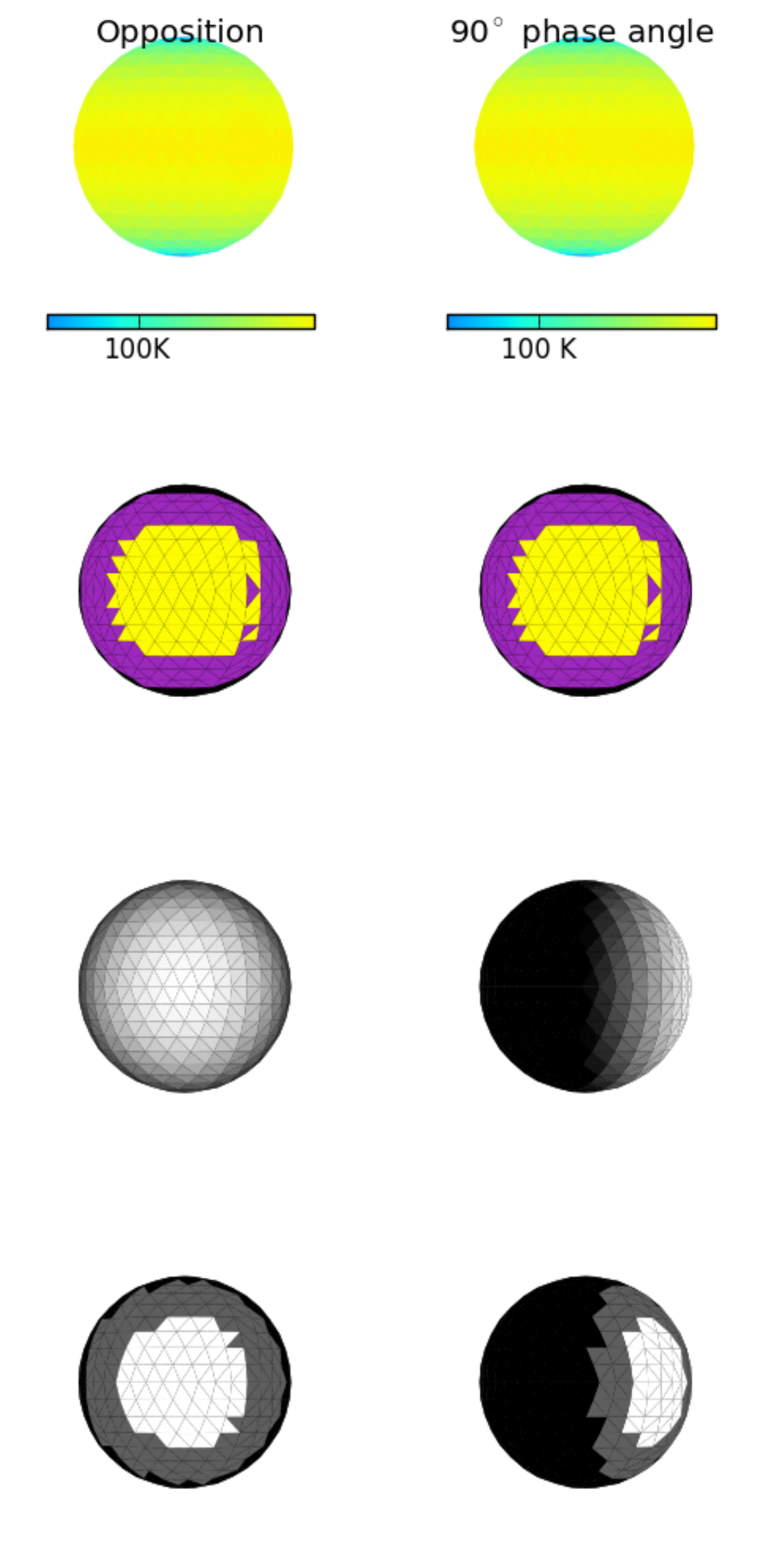}
	}
	\hfill   
    \subfloat[$\Theta = 4.2 \times 10^0$, x = 3.7]{
    \includegraphics[width=0.31\textwidth]{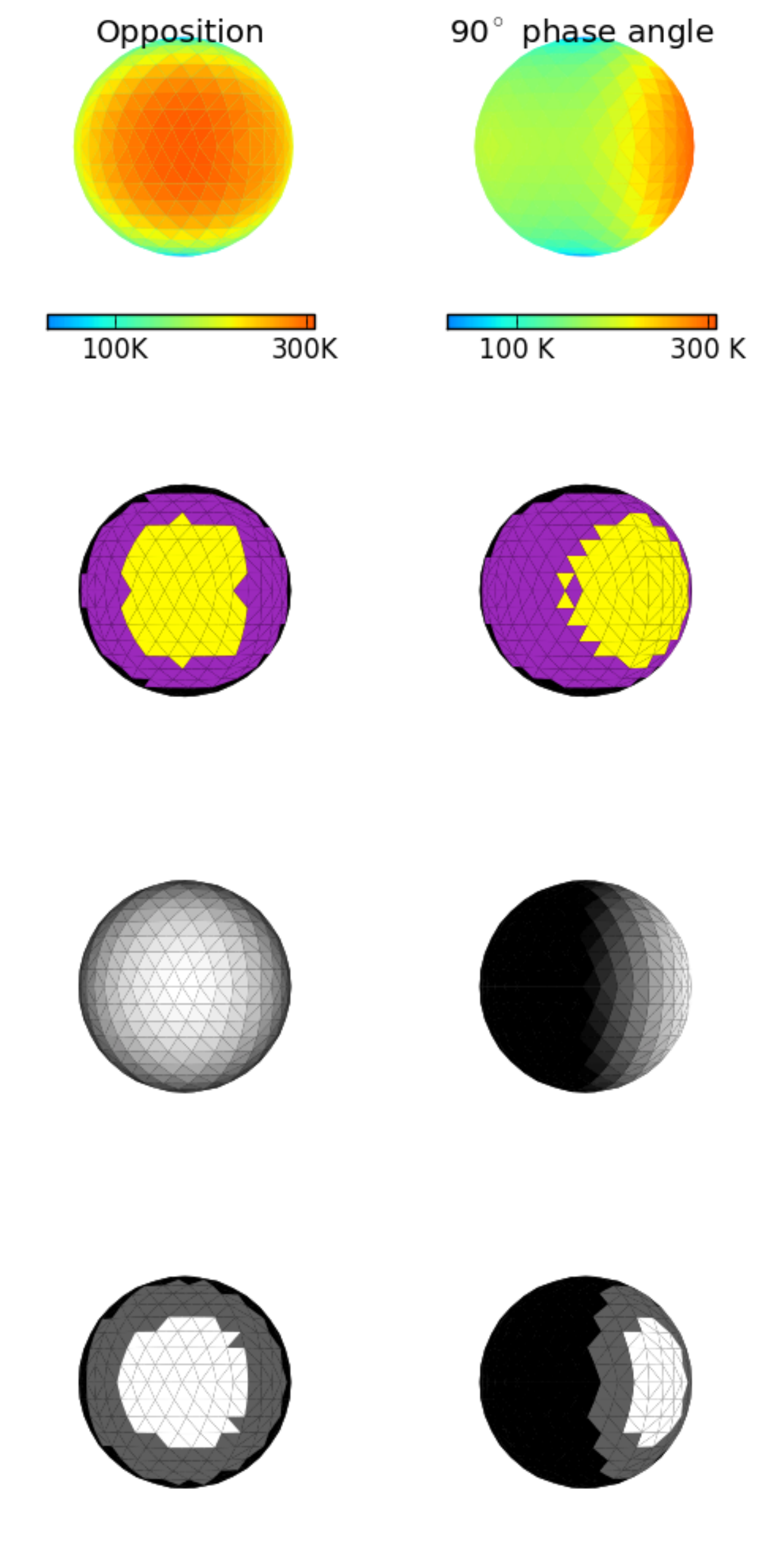}
	}
	\hfill   	
	\subfloat[$\Theta = 4.2 \times 10^{-1}$, x = 3.4]{
    \includegraphics[width=0.31\textwidth]{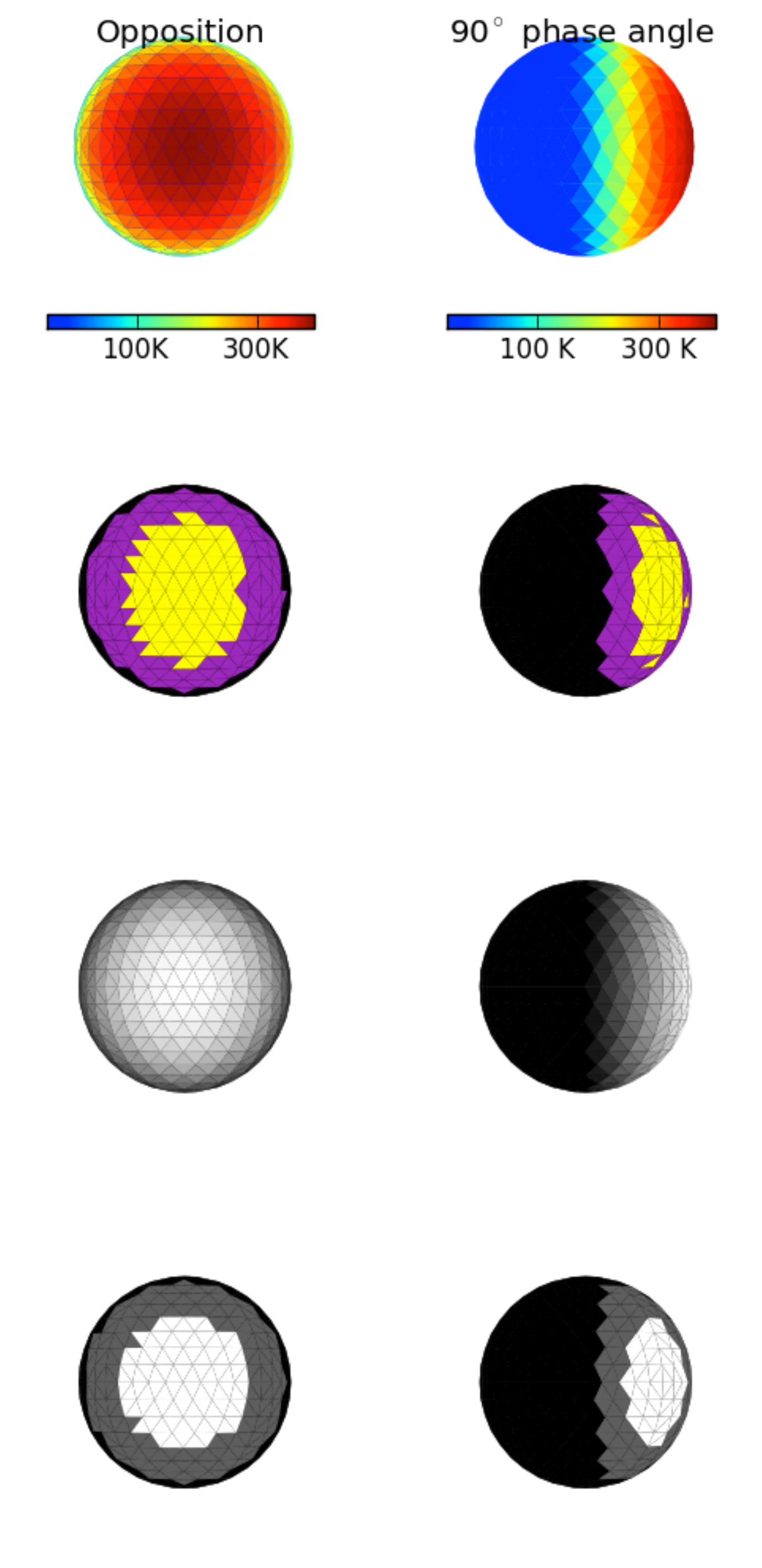}
	}
    \caption{Observed surface area of spherical asteroid in the visible and infrared for various values of $\Theta$ (x values, which are described in the Methods section, are also given). In each subfigure, the left column shows the asteroid as observed at opposition, with the sun behind the observer. The right column shows the asteroid as observed at $90^\circ$ phase angle. The top row is a temperature map, the second row shows the area producing $68\%$ (yellow) and $95\%$ (yellow and purple) of the observed thermal flux. Third row shows reflected visible sunlight. The fourth row shows the area reflecting $68\%$ (white) and $95\%$ (white and gray) of the observed optical flux.}
    \label{fig:maps}
\end{figure}

\begin{figure}[t!]
    \subfloat[$\Theta = 4.2 \times 10^2$]{
    \includegraphics[width=0.31\textwidth]{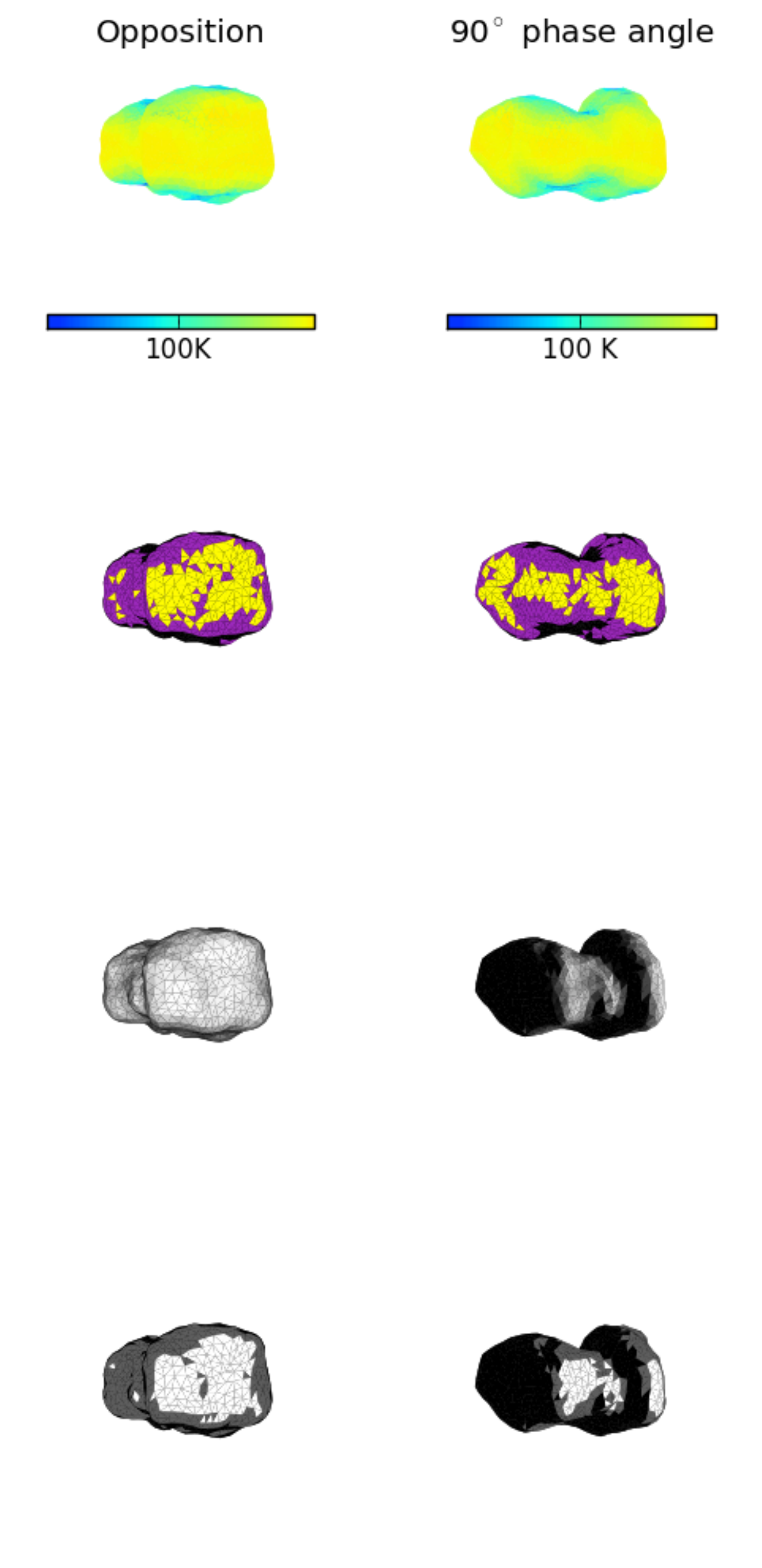}
	}
	\hfill   
    \subfloat[$\Theta = 4.2 \times 10^0$]{
    \includegraphics[width=0.31\textwidth]{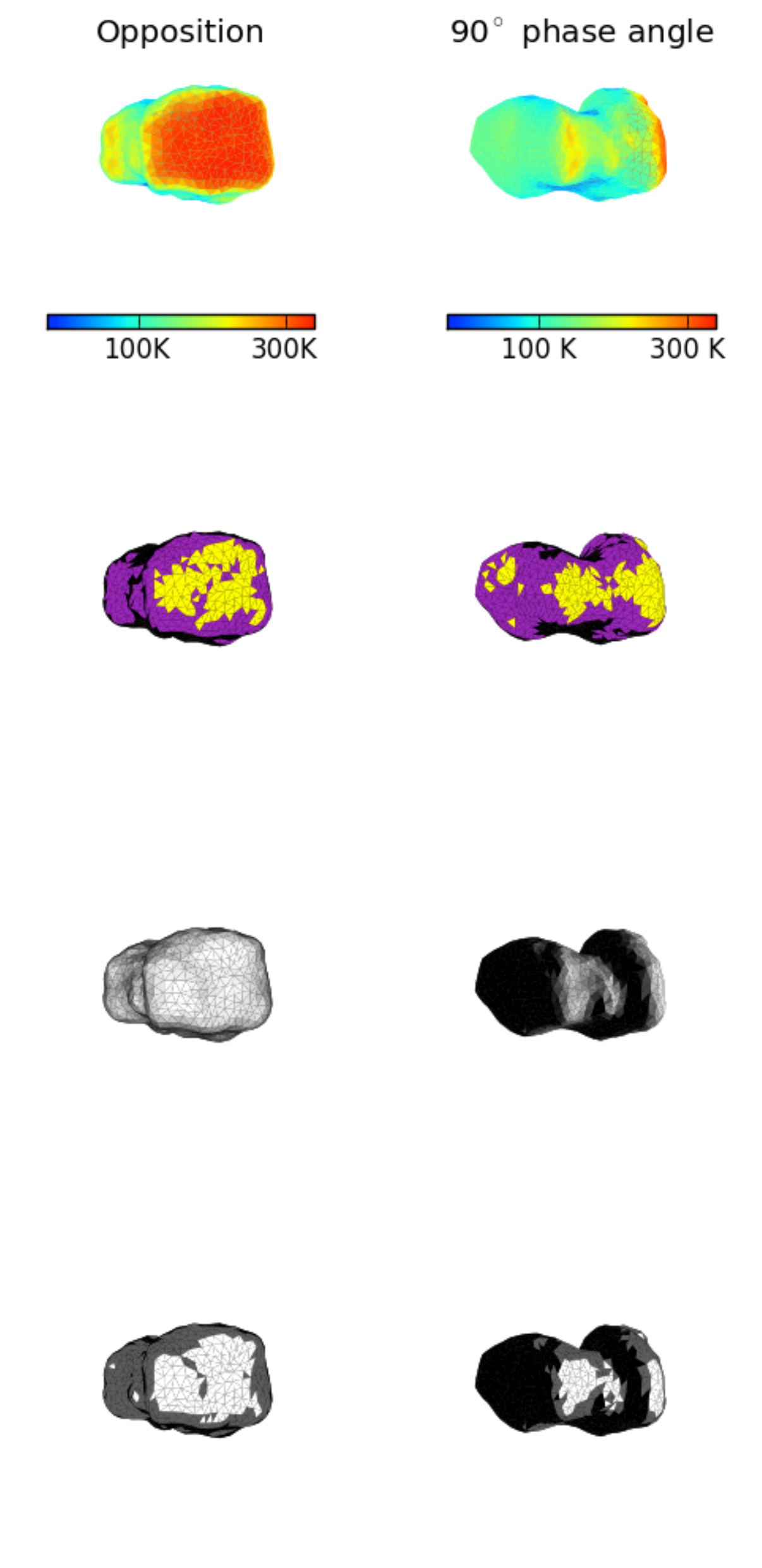}
	}
	\hfill   	
	\subfloat[$\Theta = 4.2 \times 10^{-1}$]{
    \includegraphics[width=0.31\textwidth]{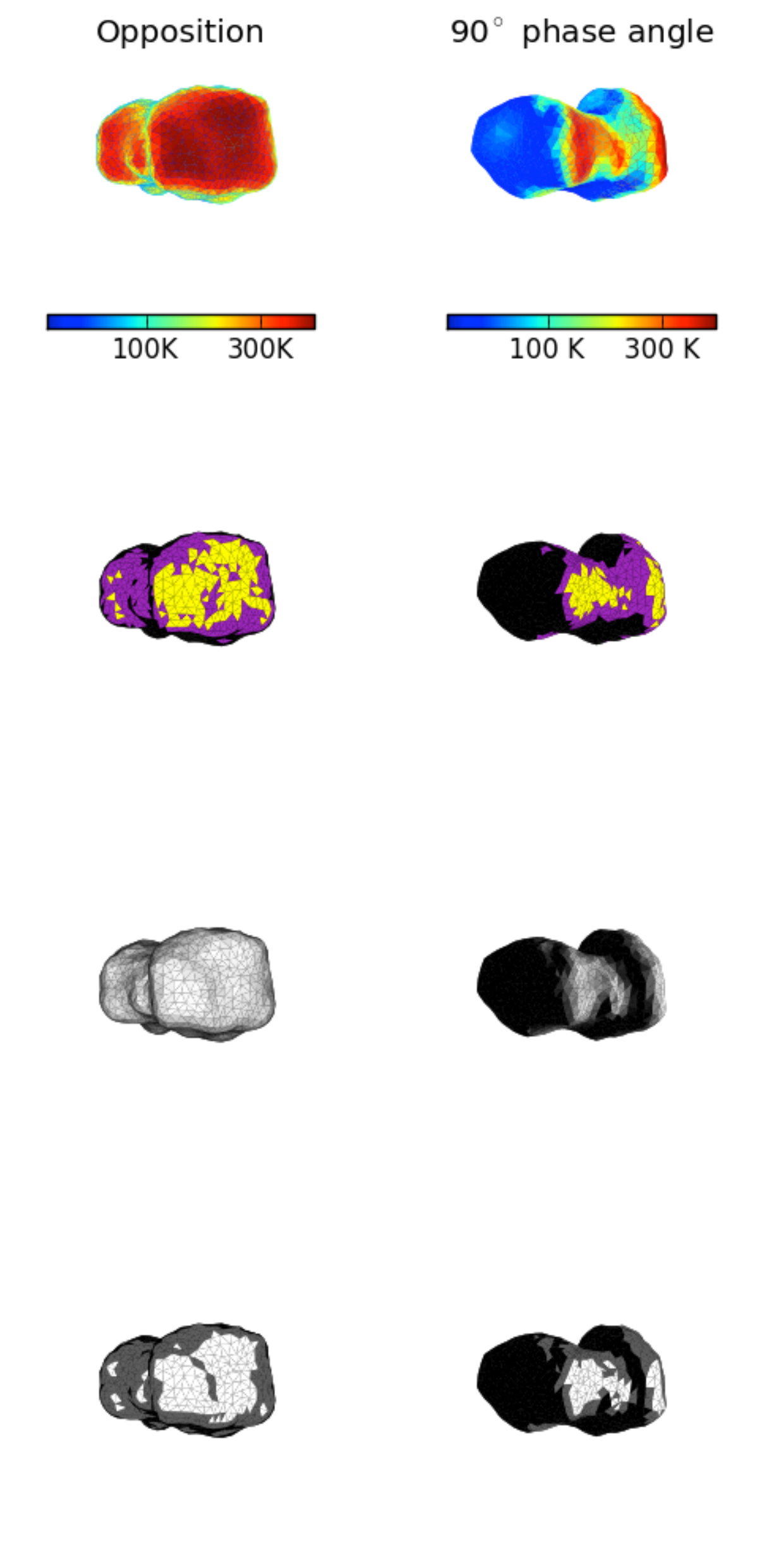}
	}
    \caption{Observed surface area of dumbbell shaped asteroid in the visible and infrared. Color coding is the same as Figure \ref{fig:maps}. }
    \label{fig:mithra}
\end{figure}

Maps showing the observed surface area as a function of $\Theta$ are given in Figures \ref{fig:maps} and \ref{fig:mithra}. These maps are for the case without surface roughness; maps of results including roughness are similar. Temperature maps and surface area contributing $68\%$ and $95\%$ of observed light in the thermally emitted IR at 12 $\mu$m are shown. Maps of observed reflected visible wavelength sunlight at opposition and at 90$^\circ$ phase angle illustrate geometric viewing effects for reference. The amount of reflected light depends on the cosine of the normal facet angle and the vector toward the sun as well as the cosine of the normal facet angle and the vector to the observer. Maps of observed thermally emitted light depend on both the temperature of the facet, and the cosine of the normal facet angle and the vector to the observer.

For the case of $\Theta = 4.2 \times 10^2$, temperature is uniform with longitude, with a hot equatorial band. For a spherical object, the observed area in the infrared does not change based on viewing angle.  The yellow region, which as a whole contributes to $68\%$ of the emitted thermal flux,  extends over this hot region, excluding the cold pole. This case is a classic ``fast rotator'', and could be accurately modeled using the FRM.

As $\Theta$ decreases, the sub-solar hotspot becomes more pronounced. For a sphere observed at opposition, a smaller area contributes to $68\%$ of the infrared flux as $\Theta$ decreases. Observed at 90$^\circ$ phase angle, a slightly larger fraction of the surface is observed in the IR vs the visible when $\Theta = 4.2 \times 10^0$, though overall flux will be lower due to colder observed area. When $\Theta = 4.2 \times 10^{-1}$, the nightside emission is effectively zero, and the temperature distribution resembles NEATM model. 

For a dumbbell shaped asteroid (Figure \ref{fig:mithra}), the irregular surface of this object means that the observed surface area in the infrared (yellow) is fragmented, indicating sensitivity to facet orientation. This fragmentation is apparent at all values of $\Theta$, particularly when compared to the observed area in visible wavelengths (reflected light, white). Sub-figure \ref{fig:mithra}(c) illustrates a low $\Theta$ or slow rotating case. When observed at $90^\circ$ phase angle, a small fraction of the surface produces $68\%$ of the observed infrared flux.  Thermal flux is not only sensitive to the orientation of each facet, but it is dependent on only a small fraction of facets.

The fraction of observed surface area in the IR varies as a function of $\Theta$ and wavelength. Observing the spherical object at opposition, $25\%$ to $29\%$ of the surface contributes $68\%$ of the observed flux at $12 \mu m$. For the dumbbell shaped object, only $12\%$ to $19\%$ of the surface contributes $68\%$ of the observed flux at $12 \mu m$. The fraction of observed surface is not sensitive to roughness when viewed at opposition, however when observed at 90 degrees phase angle, observed area decreases (by $\sim1\% -\sim3\%$) as $\epsilon_{IR_{eff}}$ decreases.

\begin{figure}[t!]
	\subfloat[Sphere]{
	\includegraphics[width=0.5\textwidth]{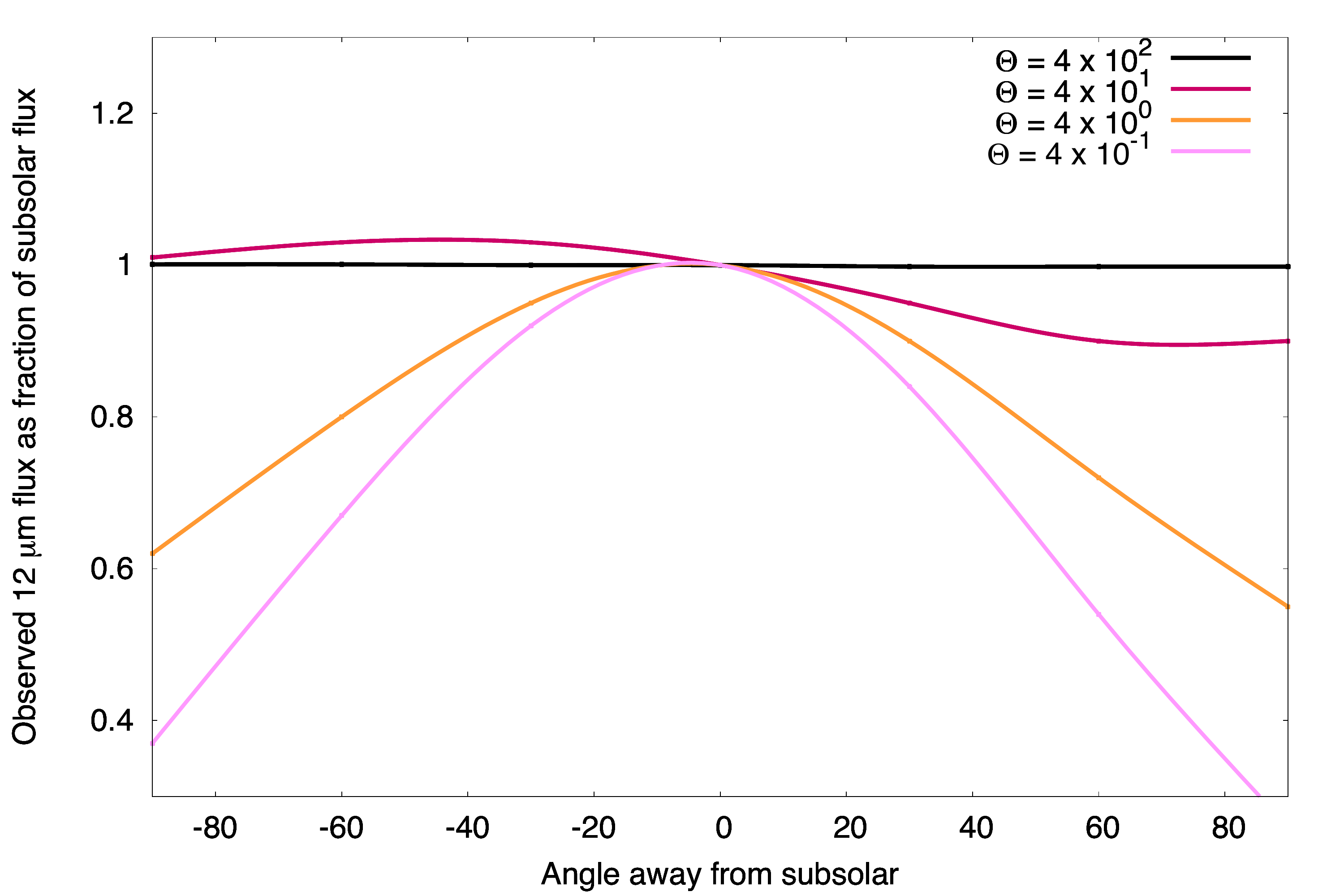}
	}
	\hfill
	\subfloat[Dumbell shape]{
	\includegraphics[width=0.5\textwidth]{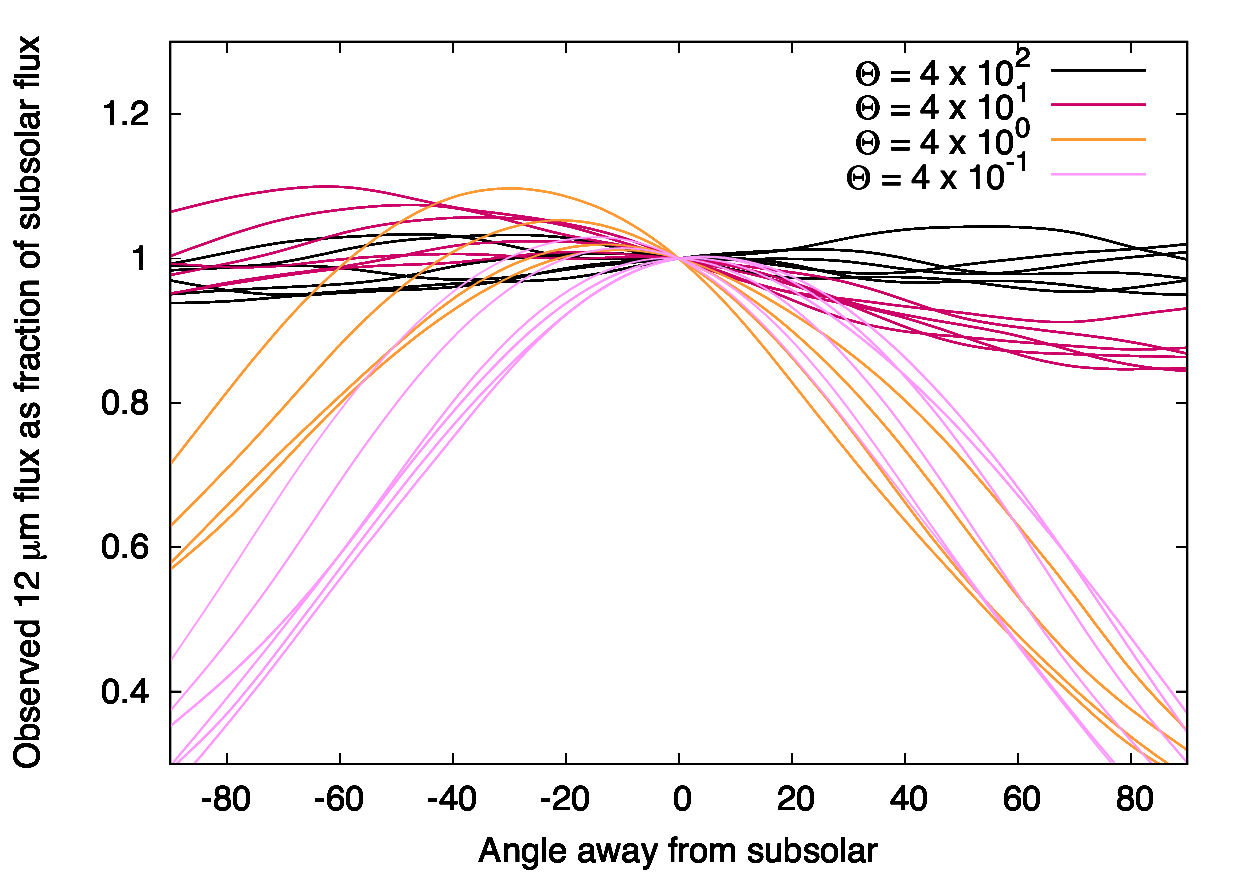}
	}
	\caption{
	Observed $12$ $\mu m$ flux as a fraction of sub-solar flux as a function of angle away from the sub solar point. (a) Spherical object, case without roughness}. Thermal lag is only present for a limited range of $\Theta$. For $\Theta=4 \times 10^1$ afternoon flux excess is $ < 5\%$ of the flux at the sub solar point, while morning flux deficit is $\sim 10\%$ of the flux at the sub solar point. (b) Dumbbell shaped object, case without roughness. Since the observed W3 flux varies with rotation for this object, five lines for each value of $\Theta$ are plotted, each assuming a different rotation phase.
	
  \label{fig:lag}
\end{figure}

Figure \ref{fig:lag} illustrates the behavior of the hot spot on the asteroids. For a spherical asteroid in the case of high $\Theta$, temperature is uniform with respect to longitude, and no hot spot exists. For $\Theta$ = $4.2 \times 10^1$, a hot spot exists, and it is significantly shifted away from $0^\circ$ due to thermal lag. This shift is responsible for powering the diurnal Yarkovsky effect \citep{2006AREPS}. For cases of $\Theta \la 4.2 \times 10^1$, the asteroid is heating and cooling quickly relative to its rotation rate, and the hotspot is located close to the sub solar point. 

The technique of determining an asteroid's spin sense (prograde vs retrograde) based on comparison of flux emitting from each side of the sub solar point (morning vs. afternoon observations) therefore is limited to a range of $\Theta$ of order 10 for this simple case, and may exclude slow rotating bodies with highly thermally insulating surfaces, fast spinners, or metallic objects with efficient thermal conduction. Additionally, even for optimal values of $\Theta$, afternoon flux excess is $ < 5\%$ of the flux at the sub solar point, while morning flux deficit is $\sim 10\%$ of the flux at the sub solar point. Implementation of this technique requires observations with sufficiently high signal to noise.

The location of the hot spot on the surface of the asteroid is related to the diurnal Yarkovsky effect. Although Figure \ref{fig:lag} illustrates how displacement is largest  on spherical asteroids with $\Theta$ of order 10, this figure is also normalized to sub solar flux. For predictions of thermal force as a function of $\Theta$, see \citet{VokDiurnalYark}, which predicts a peak when $\Theta = 1.55$.

For the dumbbell shaped asteroid, shape effects introduce complications. For a fast-rotating object ($\Theta$ = $4.2 \times 10^2$), an increase of thermal flux compared to the sub solar point can be observed on either side of the sub-solar point, depending on the rotational phase of the asteroid. Therefore, comparison of morning vs afternoon thermal flux for a fast rotator cannot indicate rotation sense. Although this is not a surprising result, fast-rotating cases must be identified and excluded from samples before a rotation sense analysis can be conducted. For this case, rotation sense could be determined during some, but not all, rotational phases when $\Theta \la 4.2 \times 10^1$, given observations with sufficiently high signal to noise.

\section{Conclusions}

When solving for the temperature of an asteroid during thermophysical modeling, an observed area should also be calculated. This is a computationally simple step that ensures that results may be interpreted accurately. Published radar-based shape models have figures that illustrate which part of the derived shape was fit to the radar data, and which parts were filled in by the inversion software. Since thermophysical modeling is sensitive to the orientation of each facet, best results will be produced when the observed area in the infrared corresponds with the area of the shape that was observed by radar. Additionally, for thermophysical modeling, shape models can be effectively employed only if their spatial resolution is smaller than the area observed in the thermal infrared, which varies with $\Theta$. 

Although telescopic images in the IR are not disk-resolved, they can be paired with an asteroid shape model of sufficiently high resolution to determine what areas of the asteroid are producing the flux observed. For example, case (c) in Figure \ref{fig:mithra} shows that only the limb of the asteroid is imaged in the infrared at $90^\circ$ phase angle. If an infrared light curve with sufficiently high signal to noise was obtained for this object, shape effects could be removed, and thermal inertia could be solved for at each of the different rotation phases. Combined with observed surface maps, the thermal inertias for each rotation phase could produce a rough map of thermal inertia.

NEATM and FRM have been used to return reliable diameters and albedos. For values of $\Theta$ comparable with the majority of asteroids, they produce temperature distributions that closely match thermophysically-derived distributions. Diameters have been verified by independent methods such as radar and stellar occultations, e.g. \cite{Mainzer2011c, Mainzer2011d,Masiero11, Masiero12, Nugent15, Nugent16a}. Although this type of modeling contains simplifying assumptions, the impacts of these assumptions are accounted for by the associated statistical uncertainties obtained when averaging over many objects ($\pm10\%$ to $\pm20\%$ $1-\sigma$ in diameter in the literature referenced previously, depending on data used). As the number of asteroids observed by radar has been rapidly growing, future work will compare NEATM-derived diameters to radar-derived diameters using a larger sample.

\section{Acknowledgments}
We thank the referee for their thoughtful comments that strengthened the manuscript.

We also thank Dr. Mark Lysek and Dr. John Schiermeier of the Jet Propulsion Laboratory, and Matt Garett of CRTech for their help with adapting the SINDA code to this work. 

We acknowledge the support of NEOWISE for the computing facilities used in this research.

\bibliographystyle{apj_cn}

\end{document}